# Sequestration of Malevolent Anchor Nodes in Wireless Sensor Networks using Mahalanobis Distance

Jeril Kuriakose[1], V. Amruth[2], Swathy Nandhini[3] and V. Abhilash[4]

[1]*School of Computing and Information Technology (SCIT), Manipal University Jaipur, Jaipur, India*
[2]*Department of Information Science and Engineering, Bearys Institute of Technology, Mangalore, India*
[3]*Department of Information Technology, Jayam College of Engineering and Technology, Dharmapuri, India*
[4]*Freelancer*
e-mail: [1]*jeril@muj.manipal.edu*

**Abstract.** Discovering the malicious or vulnerable anchor node is an essential problem in wireless sensor networks (WSNs). In wireless sensor networks, anchor nodes are the nodes that know its current location. Neighbouring nodes or non-anchor nodes calculate its location coordinate (or location reference) with the help of anchor nodes. Ingenuous localization is not possible in the presence of a cheating anchor node or a cheating node. Nowadays, it's a challenging task to identify the cheating anchor node or cheating node in a network. Even after finding out the location of the cheating anchor node, there is no assurance, that the identified node is legitimate or not. This paper aims to localize the cheating anchor nodes using trilateration algorithm and later associate it with Mahalanobis distance to obtain maximum accuracy in detecting malicious or cheating anchor nodes during localization. We were able to attain a considerable reduction in the error achieved during localization. For implementation purpose, we simulated our scheme using ns-3 network simulator.

*Keywords:* Mahalanobis distance, Trilateration, Anchor node, security, Distance-based localization, Wireless sensor networks.

## 1. Introduction

Wireless ad hoc and sensor networks are on a steady rise in the recent decade. This is because of their reduced cost in deployment and maintenance. Advancements in radio frequency spectrum also carved way for the improvement in the data rate for communication. Many devices belong to wireless ad hoc and sensor networks; one among them is anchor node [1,2]. Anchor nodes are the nodes that know its current location. Neighbouring nodes or non-anchor nodes calculate its location coordinate (or location reference) with the help of anchor nodes, and its working is quite referable to Light House.

The location of the nodes plays a significant role in many areas as routing, surveillance and monitoring, and military. The sensor nodes must know their location reference to carryout location-based routing (LR) [3]. To find out the shortest route, the location aided routing (LAR) [4] makes use of the locality reference of the sensor nodes. In some industries the sensor nodes are used to identify minute changes as pressure, temperature and gas leak, and in military, robots are used to detect land mines where in both the cases location information plays a key part.

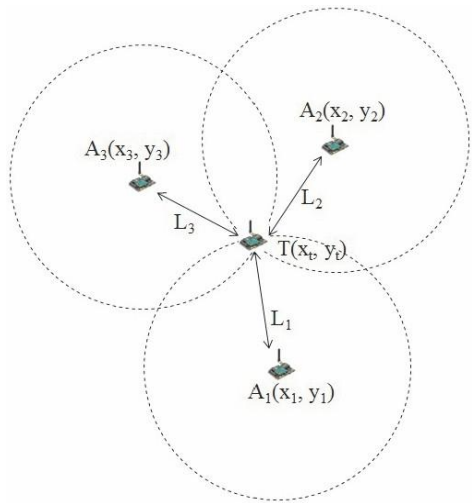
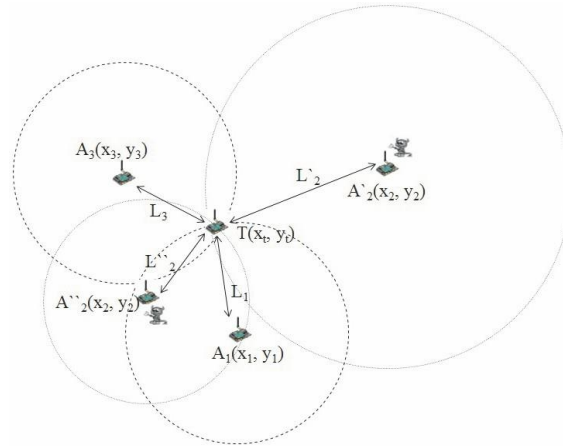

Figure 1.  Initial setup of anchor nodes.    Figure 2.  Anchor nodes after attack.

Anchor nodes can also be used to find the current location of any device (mobile phones, objects and people). It does that by transmitting anchor frames periodically or at regular intervals. Usually anchor frames are used to advertise the occurrence of a wireless modem or an Access Point (AP). Each anchor frame carries some details about the configuration of AP and a little security information for the clients.

When the technologies are on a massive upswing, the need for security of the relevant technologies arises. There can be several occasion where the anchor nodes can be vulnerable to security breach. Because of the security breach the anchor node starts cheating by providing false information. In the presence of cheating anchor nodes the odds of localization drastically decreases. Many papers [5–8] discuss about the localization of cheating anchor nodes, but with inconsistent accuracy. So, to overcome this, we localize the cheating or vulnerable anchor node using trilateration technique and associate the results with Mahalanobis distance [9,10].

**Organization of the paper** Section 2 provides the localization using trilateration algorithm and section 3 studies the Mahalanobis distance. Simulation and results are covered in section 4 and section 5 concludes the paper.

## 2. Localization using Trilateration Algorithm

Anchor nodes are widely used for tracking and localization; whereas nowadays it is also used for navigation and route-identification. With the help of anchor nodes, a user can find out his current location. Consider a scenario like a hotel or museum, there may be many occasions where people go out of track. This can be flabbergasted by mounting anchor nodes in various locations, so that people can trace out there location very easily and it is possible only when the anchor nodes are authentic. Nowadays hackers are on a rise; anybody can easily get into any system and change its settings. Similarly, they can hack any anchor nodes and change its location reference to some other false location reference, making people lose their track; thus leading to a bad imprint about the system (i.e., hotel, museum).

An attack is exemplified in Figure 1 and Figure 2. Figure 1 shows the initial deployment of anchor nodes $A_1, A_2, A_3$; with location reference $(x_1, y_1), (x_2, y_2), (x_3, y_3)$; and distance $L_1, L_2, L_3$; respectively, from the trilateration point $T$, having location reference $(x_t, y_t)$. Figure 2 demonstrates the logical deployment of anchor nodes after the attack i.e., multiple changes in location reference of anchor node $A_2$.

The three dimensional location coordinate of any device or node can be estimated using trilateration calculations. Trilateration technique uses distance measurements rather than angular measurements; the latter technique is also used in many localization techniques [11]. Using some iterative schemes like least square, least median square [6], least trimmed square [12] and gradient descent [13], can equitably increase the accuracy of trilateration technique.

Trilateration techniques use the distance measurement between the nodes to calculate the location reference. The distances between the nodes are identified using Received Signal Strength (RSSI) [11] or Time of Arrival (ToA) [11] or Time Difference of Arrival (TDoA) [11]. When a node (requesting node) wants to identify its location information using trilateration technique, it does with the help of three or more neighbouring anchor nodes.

The mathematical computation of trilateration is as follows: Consider three circles or spheres with centre $C_1$, $C_2$ and $C_3$, radius $L_1$, $L_2$ and $L_3$ from points $A_1$, $A_2$ and $A_3$ (anchor node location), refer Figure 3. The general equation of the sphere is

$$\sum_{k=1}^{3}(A_k - C_k)^2 = L^2$$

This can be modified as follows,

$$L_1^2 = A_1^2 + A_2^2 + A_3^2 \tag{1}$$

$$L_2^2 = (A_1 - D)^2 + A_2^2 + A_3^2 \tag{2}$$

$$L_3^2 = (A_1 - i)^2 + (A_2 - j)^2 + A_3^2 \tag{3}$$

Subtracting equation (2) from equation (1), we get

$$L_2^2 - L_1^2 = (A_1 - D)^2 + A_2^2 + A_3^2 - A_1^2 - A_2^2 - A_3^2 \tag{4}$$

Substituting we get,

$$A_1 = \frac{L_1^2 - L_2^2 + D^2}{2D} \tag{5}$$

From the first two circles we can find out that the two circles intersect at two different points, that is

$$D - A_1 < A_2 < D + A_1 \tag{6}$$

Substituting equation (5) in equation (1), we can procure

$$L_1^2 = \left(\frac{L_1^2 - L_2^2 + D^2}{2D}\right)^2 + A_2^2 + A_3^2 \tag{7}$$

Substituting we get the solution of the intersection of two circles

$$A_2^2 + A_3^2 = L_1^2 - \frac{(L_1^2 - L_2^2 + D^2)^2}{4D^2} \tag{8}$$

Substituting equation (1) with equation (3) and (8), we get

$$L_3^2 = (A_1 - i)^2 + (A_2 - j)^2 + L_1^2 - A_1^2 - A_2^2 \tag{9}$$

$$A_2 = \frac{L_1^2 - L_2^2 - A_1^2 + (A_1 - i)^2 + j^2}{2j}$$

$$= \frac{L_1^2 - L_2^2 + i^2 + j^2}{2j}$$

$$A_2 = \frac{i}{j} L_1 \tag{10}$$

From equation (5) and equation (10) we get the values of $A_1$ and $A_2$ respectively. From that we can find out the value of $A_3$ from equation (1),

$$A_3 = \pm \sqrt{L_1^2 - A_1^2 - A_2^2}$$

From the above equation we can say that, $A_3$ can have either positive or negative value. If any one circle intersect the other two circles precisely at one point, then $A_3$ will get a value zero. If it intersects at two or more points, outside or inside it can get either a positive or negative value, respectively.

During deployment each node carries out the trilateration process with all of its neighbouring nodes and every node is authorized with two or more trilateration points for security reasons. Every node reveals the information about its trilateration point to its immediate or one hop neighbours. Care is taken that no node reveals the trilateration information about its neighbours.

The algorithm for setting up the anchor nodes according to trilateration are as follows:

Data: Deployment of anchor nodes
Result: Successfully deploying anchor nodes and exchanging trilateration information
Start initialization;
Deploy the anchor nodes
    Set the initial coordinates (lat & long) for each anchor node;
    Cluster anchor nodes into a set of three or more;
while *not at end of deployment* do
    Trilaterate a group of anchor nodes to a centre point (or trilateration point) and save the location reference in $M_1^*$;
    Individually trilaterate all the anchor nodes with the neighbouring group and save the location references in $M_2^*$, $M_3^*$, etc.;
    Pass trilateration information to its immediate neighbours;
end
Stop deploying anchor nodes;
($^*$ $M_1$, $M_2$, $M_3$, etc., are different memory with different location reference)

Algorithm 1: Setting up anchor node

The algorithm for finding out the malicious anchor nodes are as follows:

Data: Location coordinate of nodes
Result: Finding out the malicious anchor nodes
Start;
Trilaterate each group of anchor nodes to a centre point and save that location;
Compare the obtained location with location reference ($M_1$);
while *comparison not satisfied* do
    Trilaterate all anchor nodes (individually) of the particular group
    (which does not satisfy the above comparison) with the neighbouring group
    (using the trilateration information obtained during deployment);
    Compare the obtained results with the location references ($M_2$, $M_3$, etc.);
    if a node is suspected to be malicious
        Separate the mismatched anchors node location and save the
        new location in $M_n$
    end
end
If comparison satisfied, no cheating nodes occur;
Stop;

Algorithm 2: Finding out the malicious anchor nodes

After the comparison, the anchor nodes that does not have the same location reference or the anchor node that tends to be vulnerable is considered to be malicious or cheating node. To confirm its adversary, we compare it with Mahalanobis distance. The false location coordinates of the malicious anchor nodes, identified using trilateration technique is given to the central server or aggregation point. The central server or aggregation point will be identifying the outliers with respect to the previous data set, using the Mahalanobis distance.

## 3. Correlating with Mahalanobis Distance

Mahalanobis distance applies posterior probability to identify the outliers. When two anchor nodes in space are demarcated by two or more associated location coordinates, Mahalanobis distance can be used to find the distance measure between the two anchor nodes. Mahalanobis distance identifies the malicious cheating nodes by comparing the location coordinates with respect to a centroid value. In our case the centroid value is the location coordinate of the trilateration point. The Mahalanobis distance function to identify the distance measure between two anchor nodes are as follows:

$$d(mahalanobis) = \overline{[(x_j, y_j) - (x_i, y_i)]^T * C^{-1} * [(x_j, y_j) - (x_i, y_i)]}$$

where: $d(mahalanobis)$ is the distance between two anchor nodes, $(x_i, y_i) \& (x_j, y_j)$ are the location coordinates of the two anchor nodes, $C$ is the sample covariance matrix.

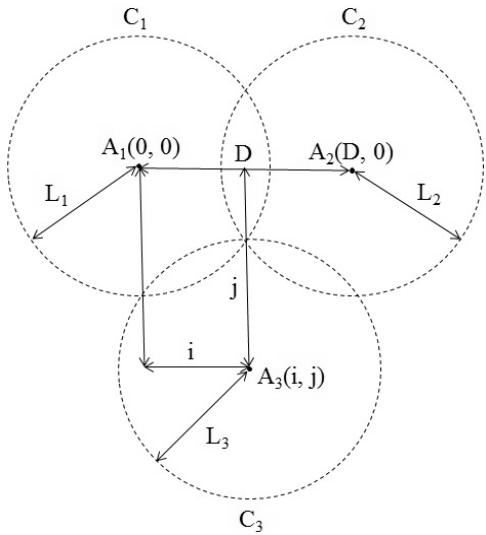

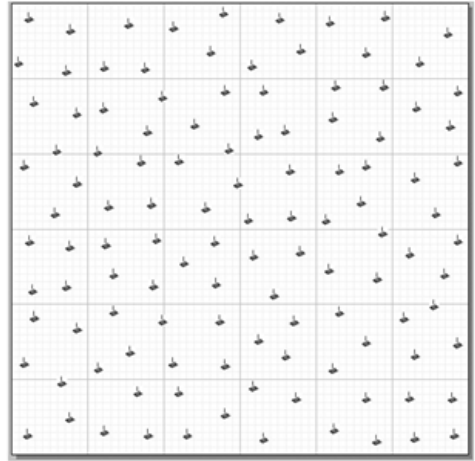

Figure 3.  Mean error in location discovery.     Figure 4.  Average time for simulation.

The variance-covariance matrix $C$ is constructed in order to gauge Mahalanobis distance,

$$C = \frac{1}{(n-1)}(x, y)^T(x, y)$$

where: $(x, y)$ is the matrix containing the location coordinates, $n$ is the number of nodes.

In the instance of multiple location references the variance-covariance will become as follows:

$$C = \begin{matrix} \sigma_1^2(x_i, y_i) & \rho_{12}\sigma_1(x_i, y_i)\sigma_2(x_j, y_j) \\ \rho_{12}\sigma_1(x_i, y_i)\sigma_2(x_j, y_j) & \sigma_2^2(x_j, y_j) \end{matrix}$$

where: $\sigma_1^2$ & $\sigma_2^2$ are the variances of the multiple location references, $\rho_{12}\sigma_1(x_i, y_i)\sigma_2(x_j, y_j)$ is the covariance between the multiple location references.

The value of $C^{-1}$ is computed as follows:

$$C^{-1} = \begin{bmatrix} \frac{\sigma_1^2(x_i, y_i)}{|C|} & \frac{-\rho_{12}\sigma_1(x_i, y_i)\sigma_2(x_j, y_j)}{|C|} \\ \frac{-\rho_{12}\sigma_1(x_i, y_i)\sigma_2(x_j, y_j)}{|C|} & \frac{\sigma_2^2(x_j, y_j)}{|C|} \end{bmatrix}$$

where: $|C|$ is the variance-covariance matrix's determinant and is equal to $\sigma_1^2\sigma_2^2(1 - \rho_{12}^2)$.

The transformation of the location coordinates to matrix form is shown in Figure 5. The Mahalanobis distance function can be modified to identify the distances from multiple location coordinates to a centroid, as follows:

$$d(\delta) = \overline{[(x_i, y_i) - (x_i, y_i)]^T * C^{-1} * [(x_i, y_i) - (x_c, y_c)]} \quad \text{for } i = 1, 2, 3, \ldots, n$$

where: $d(\delta)$ is the distance between centroid and $i^{th}$ anchor node, $(x_i, y_i)$ is the location coordinate of the $i^{th}$ anchor node, $(x_i, y_i)$ is the location coordinate of the centroid or trilateration point.

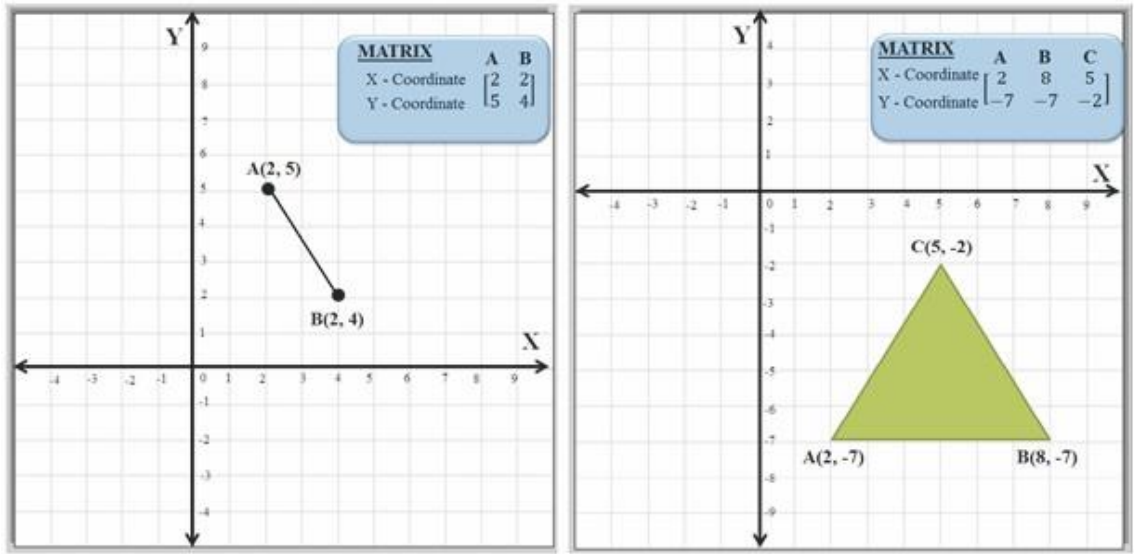

Figure 5. Transforming location coordinates to matrix representation.

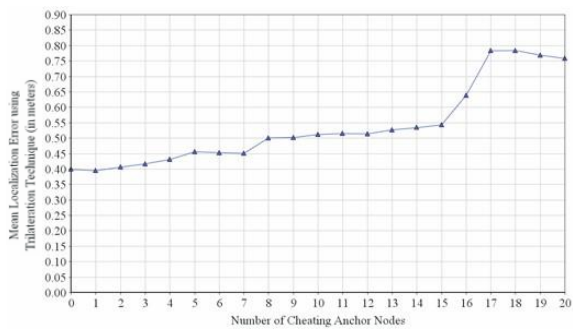

Figure 6. Mean error in location discovery.

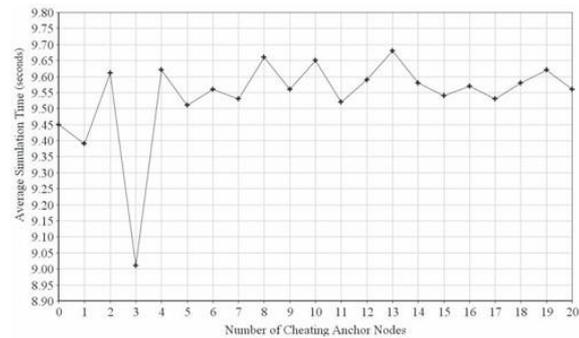

Figure 7. Average time for simulation.

The new distances obtained using Mahalanobis distance, are compared using posterior probability; leading to the confirmation of the anchor nodes adversity. The distances obtained using this method is marginally accurate than the previous method.

4. Simulations and Results

Our simulation was carried out in 600 m * 600m two dimensional environment. Deploying the anchor node accurately is very important. First three anchor nodes were placed randomly and the trilateration point is found. An anchor node is placed on the trilateration point. Any one of the first three nodes is selected and it acts as the trilateration point of the newer nodes that are going to be deployed. The above process is repeated until the final node is deployed. We deployed around 122 nodes (around 1 node for every 5 m ∗ 5 m), spread randomly from the above method. Figure 4 shows the deployment of the nodes in our scenario.

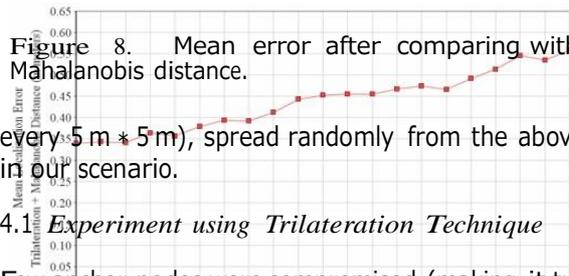

Figure 8. Mean error after comparing with Mahalanobis distance.

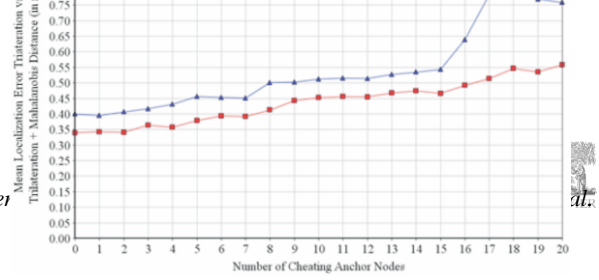

Figure 9. Result comparison.

4.1 *Experiment using Trilateration Technique*

Few anchor nodes were compromised (making it transmit false information regarding its current location) randomly and the malicious anchor nodes were found out using trilateration technique. The localization error transpired while localizing the malicious anchor nodes from random samples, were noted down. Figure 6 shows the error in location discovery and Figure 7 shows the time taken to locate the malicious anchor nodes during simulation.

4.2 *Comparing with Mahalanobis Distance*

The central server or aggregation point has a list of initial location references of the anchor nodes. The false location of the malicious anchor nodes obtained, were compared with the results obtained from Mahalanobis distance. Comparing the results obtained, reduced the error in location discovery. Figure 8 shows the mean error in locating malicious anchor nodes. Figure 9 shows the comparison of the two results, trilateration and trilateration with Mahalanobis distance. Finally the information about the malicious anchor node is conveyed to all the nodes other than the infected nodes, and the routing table is updated by impounding the malicious anchor node.

5. Conclusion

In this paper we discussed about localizing malicious anchor nodes using trilateration technique and compared the results obtained with Mahalanobis distance method. This way we were able to reduce the error attained during localization. Using Mahalanobis distance method we can obtain consistent and proficient results. Our results show that as the malicious anchor nodes increases, the simulation time and error obtained during location discovery slightly increases. The accuracy obtained in our work can be used as assistance in some wireless applications.

References

[1] P. Bahl and V. N. Padmanabhan., "RADAR: An In-Building RF Based User Location and Tracking System," *Proceedings of IEEE INFOCOM*, (2000).